\documentclass[10pt,journal]{IEEEtran}
\IEEEoverridecommandlockouts
\usepackage{cite}
\usepackage{todonotes}
\usepackage{amsmath,amssymb,amsfonts}
\usepackage{algorithmic}
\usepackage{graphicx}
\usepackage{textcomp}
\usepackage{xcolor}
\usepackage{multirow}
\usepackage{epstopdf}
\usepackage{booktabs}
\usepackage{bigstrut}
\def\BibTeX{{\rm B\kern-.05em{\sc i\kern-.025em b}\kern-.08em
    T\kern-.1667em\lower.7ex\hbox{E}\kern-.125emX}}

\usepackage{url}
\makeatletter
\def\bstctlcite{\@ifnextchar[{\@bstctlcite}{\@bstctlcite[@auxout]}}
\def\@bstctlcite[#1]#2{\@bsphack
  \@for\@citeb:=#2\do{%
    \edef\@citeb{\expandafter\@firstofone\@citeb}%
    \if@filesw\immediate\write\csname #1\endcsname{\string\citation{\@citeb}}\fi}%
  \@esphack}
\makeatother 

\usepackage{fancyhdr}

\begin{document}

% HEAD para el post-print

\pagestyle{fancy}
\fancyhf{} % sets both header and footer to nothing
\renewcommand{\headrulewidth}{0pt}
\fancyhead{} % clear all header fields
\fancyhead[C]{\fontsize{8}{10} \selectfont This article has been accepted for publication in IEEE Conference on Standards for Communications and Networking (CSCN), 2019. }

%\title{Analysis of key aspects in 3GPP/NFV standards for sharing gNB components between RAN slice subnets} % OPCIÓN ÓSCAR "MALA"

%\title{3GPP/NFV Standards for Sharing\\
%gNB components in RAN Slicing} % OPCIÓN JUANMA #1

%\title{3GPP/NFV Standards to Share gNB Components\\ in 5G RAN Slicing} % OPCIÓN JUANMA #2 (LA ÚLTIMA QUE HABÍA PUESTO)

\title{Sharing gNB components in RAN slicing: A perspective from 3GPP/NFV standards}

\author{\IEEEauthorblockN{Oscar Adamuz-Hinojosa\IEEEauthorrefmark{1}\IEEEauthorrefmark{2}, Pablo Mu\~{n}oz\IEEEauthorrefmark{1}\IEEEauthorrefmark{2}, Pablo Ameigeiras\IEEEauthorrefmark{1}\IEEEauthorrefmark{2}, Juan M. Lopez-Soler\IEEEauthorrefmark{1}\IEEEauthorrefmark{2}}

\IEEEauthorblockA{\IEEEauthorrefmark{1}Research Center on Information and Communication Technologies, University of Granada.}\\
\IEEEauthorblockA{\IEEEauthorrefmark{2}Department of Signal Theory, Telematics and Communications, University of Granada.}\\
\IEEEauthorblockA{ Email: \{oadamuz, pabloml, pameigeiras, juanma\}@ugr.es\IEEEauthorrefmark{1}\IEEEauthorrefmark{2}}
}

% FOOTNOTE - IEEE COPYRIGHT
\makeatletter
\def\ps@IEEEtitlepagestyle{
  \def\@oddfoot{\mycopyrightnotice}
  \def\@evenfoot{}
}
\def\mycopyrightnotice{
  {\footnotesize
  \begin{minipage}{\textwidth}
  \centering
 ~\copyright~2019 IEEE. Personal use of this material is permitted. Permission from IEEE must be  obtained for all other uses, in any current or future media, \\ including reprinting/republishing this material for advertising or promotional purposes, creating new  collective works, for resale \\ or redistribution to servers or lists, or reuse of any copyrighted component of this work in other works.
\end{minipage}
  }
}

\maketitle

\begin{abstract}
To implement the next Generation NodeBs (gNBs) that are present in every Radio Access Network (RAN) slice subnet, Network Function Virtualization (NFV) enables the deployment of some of the gNB components as Virtual Networks Functions (VNFs). Deploying individual VNF instances for these components could guarantee the customization of each RAN slice subnet. However, due to the multiplicity of VNFs, the required amount of virtual resources will be greater compared to the case where a single VNF instance carries the aggregated traffic of all the RAN slice subnets. Sharing gNB components between RAN slice subnets could optimize the trade-off between customization, isolation and resource utilization. In this article, we shed light on the key aspects in the Third Generation Partnership Project (3GPP)/NFV standards for sharing gNB components. First, we identify four possible scenarios for sharing gNB components. Then, we analyze the impact of sharing on the customization level of each RAN slice subnet. Later, we determine the main factors that enable isolation between RAN slice subnets. Finally, we propose a 3GPP/NFV-based description model to define the lifecycle management of shared gNB components. 
\end{abstract}

\begin{IEEEkeywords}
3GPP, NFV, RAN slicing, sharing gNB components, description model.  
\end{IEEEkeywords}

\section{Introduction}
In the next years vertical industries may bring a wide variety of services with diverging requirements in terms of functionality and performance \cite{ArticuloJose}. To economically provide them over a common wireless network infrastructure, RAN slicing has emerged as a solution \cite{RANSlicingForVerticals}. It consists of the provision of multiple RAN slice subnets, each adapted to the requirements of a specific service. To that end, RAN slicing could relies on Network Function Virtualization (NFV). This technology enables the customization of the next Generation NodeBs (gNBs) present in every RAN slice subnet through their implementation as Virtualized Network Functions (VNFs).

The Third Generation Partnership Project (3GPP) is playing a significant role on RAN slicing standardization. It has defined the gNB components, i.e. the Centralized Unit (CU), the Distributed Units (DUs), and the Radio Units (RUs); and the fifth generation (5G) radio protocol stack \cite{3GPP:38401}. It has also specified the management entities and their mechanisms to handle the lifecycle of RAN slice subnets\cite{3gpp:28.533}. However, these contributions are not enough to provide RAN slice subnets because the management of those gNB components implemented as VNFs goes beyond the 3GPP scope. The leading standardization body on network virtualization is the European Telecommunication Standard Institute (ETSI), specifically the NFV group, which has defined the management framework and its mechanisms to handle the lifecycle of VNFs \cite{nfv:eve012}.

Based on the 3GPP and ETSI-NFV contributions, several research projects are developing specific solutions for RAN slicing \cite{5GPPPSumarize}. The majority of these solutions assume a single VNF instance to accommodate a gNB component of a specific RAN slice subnet. This approach guarantees the customization of each RAN slice subnet, however the resource utilization can be inefficient. For example, let us assume an individual gNB component for each RAN slice subnet and a fixed resource capacity per VNF instance. Then, if the required capacity of two or more RAN slice subnets fits into one VNF instance, sharing a VNF instance will be a more efficient solution than deploying separate VNF instances.

Sharing VNF instances could involve statistical multiplexing gains on the utilization of virtual resources. However, achieving the customization level required by each RAN slice subnet is a challenge. Some research projects such as 5G-PICTURE \cite{5GPICTURE}, SliceNet \cite{SLICENET} or 5G-MoNArch \cite{5GMONARCH} has pursued a trade-off solution between customization and resource utilization. Notwithstanding, these projects have analyzed neither the impacts of sharing gNB components on the customization of each RAN slice subnet, nor the main factors that enable the isolation between RAN slice subnets. Additionally, these projects has focused on the lifecycle management from the 3GPP viewpoint, neglecting the NFV perspective.

In this paper, we shed light on the key aspects in 3GPP/NFV standards for sharing gNB components between RAN slice subnets, typically, enhanced Mobile Broadband (eMBB), ultra-Reliable Low Latency Communication (uRLLC), and massive Machine Type Communication (mMTC). To that end, we (a) identify the main scenarios for sharing gNB components; (b) analyze the impact of sharing on the customization level of each RAN slice subnet; (c) determine the main factors that enable the isolation between RAN slice subnets; and (d) propose a description model to define the lifecycle management of a shared gNB component using the 3GPP/NFV management templates.

The remainder of this article is as follow. Section \ref{sec:background} overviews the 3GPP/NFV standardization for RAN slicing. Section \ref{sec:Analysis} analyzes the key aspects and enablers for sharing gNB components. Section \ref{sec:SharedManagementTemplates} provides a 3GPP/NFV-based description model to define the lifecycle management of a shared gNB component. Finally, Section \ref{sec:Conclusions} draws the main conclusions of this work.

\section{3GPP/NFV standardization for RAN slicing}\label{sec:background}

\subsection{3GPP Next Generation RAN architecture}
The 3GPP Next Generation RAN (NG-RAN) is composed of gNBs, which provide wireless connectivity to the User Equipments (UEs) through the New Radio (NR) protocol stack \cite{3GPP:38401}. From a functional viewpoint, a gNB is composed of RUs, DUs and a CU. The functionalities of the NR protocol stack are distributed over these components in a flexible way. The RUs comprise at least the antennas and  the radio-frequency circuitry, thus they must be implemented as hardware. The remaining functionalities might be virtualized and they are split into multiple DUs and one CU. The DUs contain the low-layer functionalities whereas the CU includes the high-layer functionalities. The aim of this split is to leverage the benefits of virtualization and centralization. Additionally, the CU could be split into two entities, each gathering the control and data plane functionalities, respectively\footnote{The benefits and drawbacks of this approach could be consulted in \cite{3gpp:38.806}}. For simplicity, this article does not assume control and data plane split.

There exists eight options to split the gNB functionalities as Fig. \ref{fig:gNBarchitecture} shows. However, there is a consensus in the industry and academia that feasible implementations in the short-term are option \#2 for CU-DU, and option \#7 for DU-RU \cite{FunctionalSplitSurvey}.

From the infrastructure perspective, the partial virtualization of gNBs requires the existence of Point of Presences (PoPs) in addition to cell sites. A PoP is a cloud site that hosts the virtual resources to accommodate VNF instances. These PoPs, classified as aggregation and edge PoPs, connect the cell sites with the core network through a hierarchical approach \cite{OscarRANSlicing}.

\subsection{Enabling RAN slicing in the NG-RAN architecture}\label{sec:NRProtocolStack}
To slice the NG-RAN architecture into multiple RAN slice subnets, the CU and the DUs should be individually deployed for each RAN slice subnet. Thereby, the gNB functionalities could be customized to meet the specific requirements of each RAN slice subnet.

\begin{figure}[tb!]
\centering 
\includegraphics[width=\columnwidth]{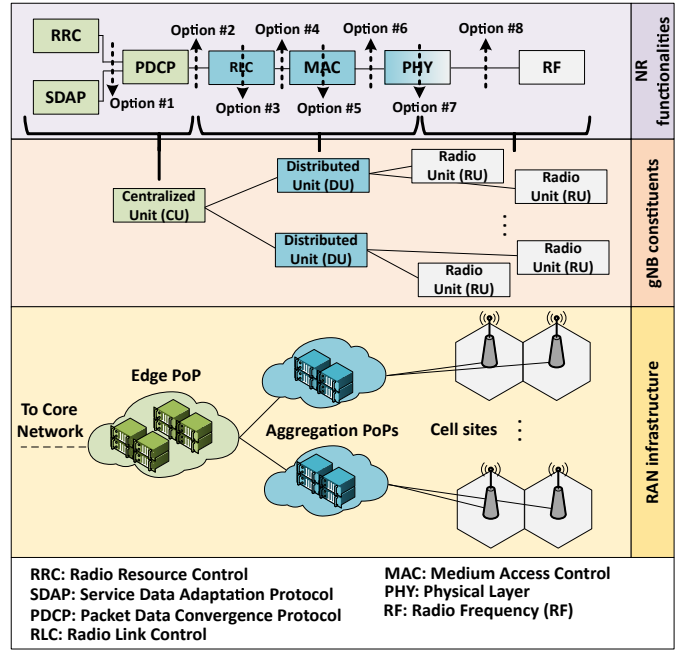}
\caption{NG-RAN architecture. For comprehensibility purposes, we assume the CU and the DUs are virtualized.}
\label{fig:gNBarchitecture}
\end{figure}

Regarding its functionalities, a gNB component comprises not only the NR functionalities depicted in Fig. \ref{fig:gNBarchitecture}, but also procedures for Radio Resource Management (RRM) \cite{RRMslicing}. Those RRM procedures that are time-sensitive, e.g., Packet Scheduling (PS), Link Adaptation (LA), etc, are hosted in the DU. The remaining RRM procedures (e.g., Mobility Management (MM), Radio Admission Control (RAC), etc) are hosted in the CU to leverage the benefits of centralization.

Each RRM procedure is controlled by a vendor-specific algorithm that guarantees the performance requirements of each UE while the available radio resources are efficiently used. To consider RAN slicing in each RRM procedure, is reasonable to implement a two-level algorithm: inter-slice and intra-slice \cite{RRMslicing}. At inter-slice level, a RRM algorithm copes with the management of all the RAN slice subnets considering the available radio resources on the whole RAN infrastructure. At intra-slice level, this algorithm is specific for a RAN slice subnet and it is designed to meet its requirements. Furthermore it only considers the allocated radio resources for this RAN slice subnet. 

To have a complete picture of the RAN infrastructure, the implementation of RRM algorithms at inter-slice level could be hosted outside the gNB. Additionally, the frequency work of the RRM algorithms (i.e., number of times that they are executed in a period of time) at inter-slice and intra-slice levels cannot be the same. The algorithm at inter-slice level deals with a enormous amount of information. This fact hinders its coordination with the intra-slice implementation of the RRM algorithm, specially if the last is time-sensitive \cite{RRMslicing}.  

Focusing on the RRM algorithms at intra-slice level, their decisions are individually applied to each Data Radio Bearer (DRB). For example, the RAC could accept the establishment request for a DRB whose user data require a specific throughput and latency. To apply the RRM decisions, the RRC layer configures the remaining layers to provide a specific treatment of the user data at each DRB \cite{3gpp:38.331}. The configured layers are: SDAP, PDCP, RLC, MAC and PHY.

\subsubsection{SDAP}
This layer is responsible for mapping the traffic flows received from the core network to DRBs with a specific Quiality of Service (QoS), thus the configuration of these DRBs is key to provide the UEs of a RAN slice subnet a service with a given QoS.

\subsubsection{PDCP}
This layer applies to each DRB functionalities such as ciphering, robust header compression, or packet duplication. The first two introduce a considerable latency, thus they could be disabled to the RAN slice subnets for uRLLC since they require low latency \cite{METISii}. On the contrary, packet duplication is recommendable for uRLLC RAN slice subnets because they require a high reliability.

\subsubsection{RLC}
This layer comprises functionalities such as segmentation and transfer mode. Concerning segmentation, only eMBB RAN slice subnets will require it to split packets on smaller units because they deal with large payloads. For transmission of user plane data, Acknowledge Mode (AM) is appropriate for URLLC RAN slice subnets. For those where reliability is not a critical requirement, Unackowledge Mode (UM) is a better option \cite{METISii}.

\begin{figure}[tb!]
\centering 
\includegraphics[width=\columnwidth]{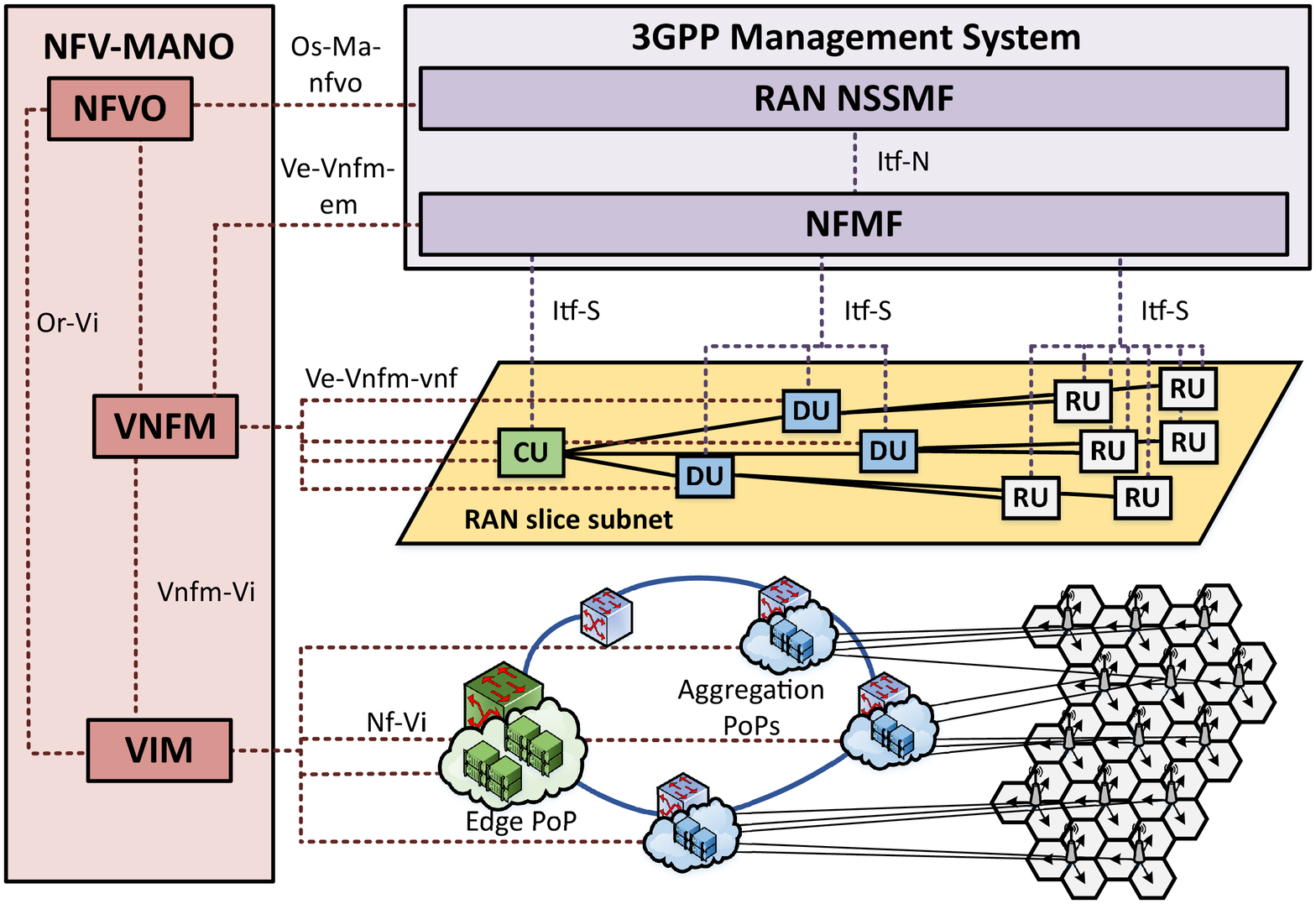}
\caption{3GPP/NFV-based framework for RAN slicing management.}
\label{fig:RANslicingframework}
\end{figure} 

\subsubsection{MAC}
This layer contains features such as the Hybrid ARQ (HARQ) or the slot format. The HARQ could be specifically configured to optimize the performance of a RAN slice subnet such as the spectral efficiency for eMBB, the coverage for mMTC or the round-trip time for uRLLC. 

For Time Division Duplex (TDD) operation, the slot format could be adapted to balance the number of OFDM symbols between the downlink and uplink as a function of the traffic symmetry in each RAN slice subnet (e.g., more downlink bits for eMBB RAN slice subnets).

Additionally, the MAC layer comprises other RRM procedures (e.g., PS, LA, etc). Focusing on PS, the algorithm and the optimization criteria could be could be adapted to optimally distribute the radio resources between the UEs attached to a specific RAN slice subnet \cite{Nikaein1}. Some examples: semi-persistent planing is better for transmitting periodic information of mMTC services; or optimization criteria such as guaranteeing latency and throughput are appropriate for uRLLC and eMBB services, respectively.

\subsubsection{PHY}
This layer is responsible for aspects such as the numerology or the Modulation and Codification Scheme (MCS). Each RAN slice subnet might require a different numerology. For example, uRLLC RAN slice subnets can benefit from higher numerologies to transmit data with lower latency due to a shorter transmission time interval. To enable a single carrier to support several numerologies, the bandwidth is divided into a set of bandwidth parts, each defining a specific numerology \cite{DaSilva1}. Thereby, if several RAN slice subnets require different numerologies, they could share the same carrier but using different bandwidth parts. In case of using the same numerology, they could also share the radio resources within a bandwidth part.

\subsection{Management of RAN slice subnets}
A RAN slice subnet comprises gNBs that are configured to provide the required behavior. In turn, the components of each gNB could be implemented as VNFs or Physical Network Functions (PNFs), i.e., dedicated hardware. 

To manage the lifecycle of RAN slice subnets, the 3GPP and ETSI-NFV have proposed in \cite{3gpp:28.533, nfv:eve012} a RAN slicing management framework as depicted in Fig. \ref{fig:RANslicingframework}. This management framework requires the interoperation of the NFV-Management and Orchestration (MANO) and the 3GPP management system. The NFV-MANO comprises three functional blocks: Virtualized Infrastructure Manager (VIM), VNF Manager (VNFM), and  NFV Orchestrator (NFVO) \cite{OscarScalado}. With these functional blocks, the RAN slicing management framework could only perform tasks related the virtualization of some gNB components, thus NFV-MANO is not enough to manage RAN slice subnets. Specifically NFV-MANO cannot (a) translate the performance and functional requirements of a gNB into the amount of the virtual resources that accommodate the gNB components; and (b) manage the Fault, Configuration, Accounting, Performance and Security (FCAPS) of the gNB components from the application perspective. These tasks are performed by the network slice subnet management service provider, and the network function service provider, both belonging to the 3GPP management system. For simplicity, we denote these entities as RAN Network Slice Subnet Management Function (NSSMF) and Network Function Management Function (NFMF), respectively (see the example of service management providers in section A.4 of \cite{3gpp:28.531}). The RAN NSSMF performs tasks (a) and (b) while the NFMF is controlled by the RAN NSSMF  to carry out the activities related to (b) in the gNB components. Since the RAN NSSMF is in charge of configuring the gNB components, it could host the inter-slice implementation of those RRM algorithms that are non-time sensitive. In such case, the RAN NSSMF would transfer the inter-slice decisions to the intra-slice algorithm implementations hosted in each gNB component.

To automate the lifecycle management of RAN slice subnets, the RAN slicing management framework relies on a set of predefined templates. The main template is the RAN Network Slice Subnet Template (NSST), proposed by the 3GPP. It could define the gNB components of a RAN slice subnet; and the parameters for its FCAPS management at application level \cite{OscarRANSlicing}. Since some of the gNB components might be virtualized, the RAN NSST must reference to the NFV management templates to describe the lifecycle of the VNFs that host them \cite{OscarRANSlicing}. In Section \ref{sec:SharedManagementTemplates}, we shed light on the utilization of the 3GPP/NFV management templates.

\section{Analysis of key aspects and enablers for sharing gNB components}\label{sec:Analysis}

\subsection{Main scenarios for sharing gNB components: Enabling customization}
\begin{figure}[t]
\centering 
\includegraphics[width=\columnwidth]{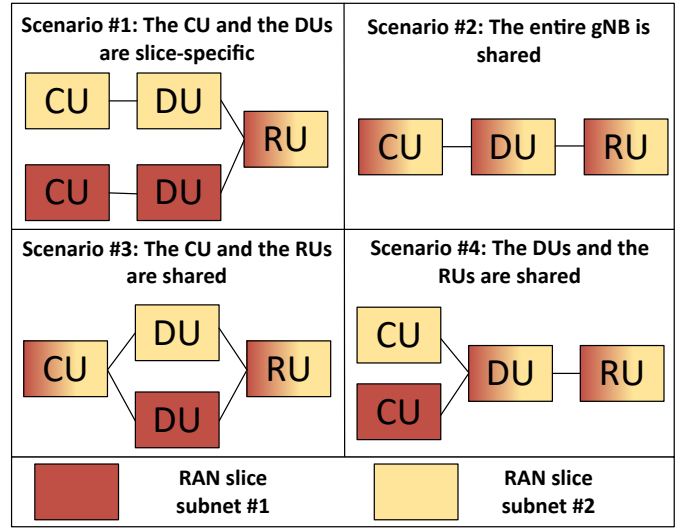}
\caption{Main scenarios for sharing the components of a gNB between several RAN slice subnets. We assume that RUs are shared in each scenario}
\label{fig:Scenarios}
\end{figure}

As depicted in Fig. \ref{fig:Scenarios}, there are four main scenarios for sharing the gNB components between several RAN slice subnets. For all scenarios, we assume the CU and the DUs to be implemented as VNFs, and the RUs as PNFs. Below, we discuss each scenario focusing on the CU and DUs.

\subsubsection{Scenario \#1. The CU/DUs are specific to each RAN slice subnet}
This scenario enables the full customization of each RAN slice subnet because the RRM algorithms at intra-slice level could be specifically implemented to meet their performance requirements (e.g., an intra-slice implementation of PS that guarantees per-UE throughput). This fact involves that, based on the decision of the intra-slice RRM algorithms in a RAN slice subnet, the RRC layer can specifically configure the DRB treatment along the entire NR protocol stack. This scenario is the easiest for earlier implementations since the gNB components are slice-agnostic.

Despite its benefits in terms of customization, this scenario is the least efficient in terms of resource utilization since it presents specific VNF instances to implement the CU/DUs of each RAN slice subnet. It might involve the underuse of the virtual resources available in the edge/aggregation PoPs. For example, let us assume a fixed resource capacity per DU instance, e.g., one virtualized CPU (vCPU), each belonging to a different RAN slice subnet. If the sum of the resource consumption of two DU instances (e.g., 65 \% and 15 \% of vCPU utilization, respectively) is less than the resource capacity of a single DU instance (e.g., 80 \% $<$ 100 \%), two vCPUs will be used when only one is required. 

A limitation of this scenario is that the isolation between RAN slice subnets might even not be guaranteed in spite of presenting separate VNF instances for their CU/DUs. For example,  implementing the VNF instances through Virtual Machines (VMs)\footnote{Note that in this article the term VM refers to a virtualization technology in general (i.e., KVM, Linux containers, dockers, etc).} hinders their isolation due to the virtualization of the Network Interface Cards (NICs) as part of the infrastructure located in an edge/aggregation PoP. Used to interconnect VMs, the virtual NIC (vNIC) of a VM could negatively affect the transmission performance on the vNICs of other VMs \cite{vNICProblem}. This is due to the fact that increasing the number of VNF instances, and thus the number of vNICs, elevates the interrupt requests and context switching time between the VMs and the hipervisor (i.e., the software, firmware or hardware that creates the VMs). This means that if the number of CU/DUs hosted in an edge/aggregation PoP is high, the RAN slice subnets that comprise these CU/DUs could suffer performance degradation even without sharing spectrum. 

\subsubsection{Scenario \#2. The entire gNB is shared between RAN slice subnets}
Unlike scenario \#1, this scenario is the most efficient in terms of resource utilization since it presents less VNF instances to accommodate the traffic demands of all the RAN slice subnets. Thereby, the utilization of the resource capacity on the aggregation/edge PoPs could be optimized. For instance, assuming the same scenario as the example used in Scenario \#1, a shared VNF instance will only require a single vCPU to process the traffic of two RAN slice subnets (i.e., a DU consuming the 80 \% of one vCPU).

In this scenario, the vNICs are shared between RAN slice subnets. This means that the user data of each RAN slice subnet transverse the same vNICs, thus the average waiting time of a packet in the vNIC buffer increases. Despite this isolation problem, the main waiting time could be reduced for higher priority packets by controlling some radio parameters in the shared gNB constituents (see Section \ref{sec:isolation} for more details). Additionally, this scenario presents a reduced number of VNF instances, and thus the number of vNICs, involving a decrease of the interrupt requests and context switching time between these instances and the hypervisor. Thereby, the transmission performance on the vNICs is not as negatively affected as in scenario \#1. 

From a functional perspective, the customization level of each RAN slice subnet could be constrained in this scenario. If the RRM algorithms at intra-slice level are shared between RAN slice subnets, the configuration of the DRB treatment in the shared NR protocol stack could not be independently adapted according to the specific requirements of each RAN slice subnet.

The solution to customize the behavior of each RAN slice subnet is making slice-aware (a) the RRM algorithms at intra-slice level and (b) the RRC layer. This means providing them the intelligence to identify the association between a RAN slice subnet and a DRB in order to specifically configure its treatment along the remaining NR protocol layers.

A key element to make slice-aware the RRM algorithms and the RRC layer is the Single Network Slice Selection Assistance Information (S-NSSAI)\cite{3gpp:23.501}. Defined by the 3GPP, this parameter classifies a network slice in one of the three main service types (i.e., eMBB, mMTC, and uRLLC). Optionally, the S-NSSAI can define a specific subtype within eMBB, mMTC, or uRLLC (e.g., for a specific vertical use case). This parameter is used for associating a Protocol Data Unit (PDU) session with a network slice. Since a PDU session comprises the QoS flows mapped to the DRBs of a specific RAN slice subnet, the slice-aware RRM algorithms (and RRC layer) could identify the association between a RAN slice subnet and a DRB through the S-NSSAI. Thereby, at intra-slice level, both the RRM algorithms and the RRC configuration can be adapted for each RAN slice subnet.

Despite the evident utility of the S-NSSAI, making slice-aware the RRM algorithms at intra-slice level (and the RRC layer) involves a higher complexity in their designs. Furthermore, the execution of these algorithms could be more costly in terms of computational requirements. This fact could degrade the performance of the time-sensitive RMM procedures located in the shared DU, thus decreasing the QoS provided by each RAN slice subnet.

Even though the RRM algorithms at intra-slice level (and RRC layer) were not slice-aware, sharing the CU/DUs could be useful for the UE attachment to several RAN slice subnets. In this use case, each RAN slice subnet would comprise two sets of gNBs, each implementing scenario \#1 and \#2, respectively. Those gNBs implementing scenario \#2 would only process signalling messages for attaching the UEs to each RAN slice subnet. After UE attachment, the user data of each RAN slice subnet would be processed by those gNBs implementing scenario \#1.

\subsubsection{Scenario \#3. The CU/RUs are shared between RAN slice subnets}

This scenario is not as efficient as scenario \#2 in terms of resource utilization because less VNF instances are shared between RAN slice subnets. However, it could present a higher level of customization since the RRM algorithms at intra-slice level located in the DUs can be specifically implemented for each RAN slice subnets.

The main drawback of this scenario is that the RRM algorithms at intra-slice level in the shared CU and the RRC layer must be slice-aware. However, the complexity of making slice-aware the RRM procedures at intra-slice level in the CU and the RRC layer is not as high as in the scenario \#2 because they are not as time-sensitive as the RRM procedures located in the DU.

Assuming slice-awareness, the shared CU should identify the DU that process the traffic of a specific RAN slice subnet. To that end, the CU could use a matching table that maps the S-NSSAI of each RAN slice subnet with the identifier of the corresponding slice-specific DU (i.e., the DU ID \cite{3GPP:38401}). Thereby, the user data associated to a DRB (specific for a RAN slice subnet) could be delivered to the correspond DU after its processing by the shared CU.

\subsubsection{Scenario \#4. The DUs/RUs are shared between RAN slice subnets}
This scenario is more efficient than the scenario \#3 in terms of resource utilization but not as efficient as the scenario \#2. The reason of that is that the number of DUs is higher than the number of CUs, thus the number of VNF instances can be considerably reduced in scenario \#4.

Regarding the level of customization, unlike the scenario \#3, the intra-slice RRM algorithms in the CU and the RRC layer are specific to each RAN slice subnet, thus their DRB treatment along the entire NR protocol can be adapted to meet their requirements. The main issue of this scenario is that the intra-slice RRM algorithms located in the DU must be slice-aware. Another drawback of this scenario is that the complexity of making slice-aware the RRM procedures is higher than the scenario \#3 because these algorithms in the DU are time-sensitive.

Assuming slice-awareness, the shared DU should identify the source/target CU for the user data associated to each DRB (specific for each RAN slice subnet) to properly process them. To that end, the DU could use a matching table that maps the identifier of each CU (i.e., CU ID) with a S-NSSAI. Thereby, extracting the CU ID of the received user data, the DU could identify the RAN slice subnet that these data belong to, and apply them a specific processing. Note that the CU ID is not currently defined by the 3GPP because assuming non-sharing scenario involves a unique CU per gNB, thus the gNB ID is enough. To enable sharing scenarios, the 3GPP should define the CU ID in future specifications.

\subsection{Sharing virtualized gNB components: Enabling isolation}\label{sec:isolation}

\begin{figure*}[tb!]
\centering 
\includegraphics[width=\textwidth]{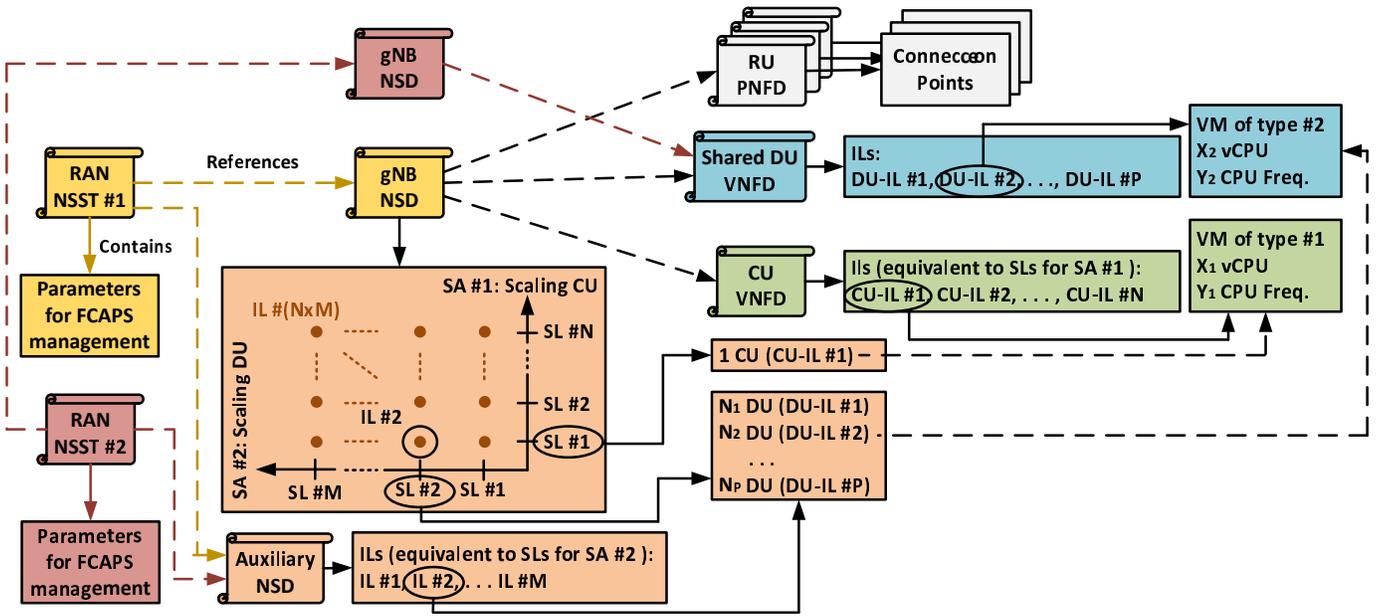}
\caption{Proposed model to describe shared DU instances using 3GPP/NFV management templates. Note that the model for sharing CU instances will be the same except for: (a) the CU VNFD is shared instead of the DU VNFD; (b) the ILs of the Auxiliary NSD would be equivalent to the SLs for SA \#1;  and (c) these ILs would reference to the CU-ILs. To avoid redundancy information, the specific CU VNFD and RU PNFDs for RAN slice subnet \#2 are not shown}
\label{fig:ProposalTemplates}
\end{figure*} 

To leverage the benefits of sharing gNB components, the isolation between RAN slice subnets must be guaranteed (i.e., if one RAN slice subnet suffers performance degradation, the performance of others RAN slice subnets must remain unaltered). This means the cumulative resource consumption of all RAN slice subnets in the VNF instances that accommodate the shared gNB components cannot exceed their resource capacity. Consequently, the processing layers in NR and operations that significantly impact the resource consumption must be controlled for each RAN slice subnet.

Several works such as \cite{DUResourceUtilization2, DUResourceUtilization3} have modeled the vCPU consumption by the PHY layer of a virtualized LTE evolved NodeB (eNB) implemented with Open Air Interface \cite{OAI}. Although they do not consider a 5G gNB, their contributions qualitatively identify the main factors that increase the vCPU consumption of a DU. These factors are: (a) a higher modulation order, which exponentially increases the CPU utilization; (b) a higher number of PRBs, which linearly increases the CPU utilization by an offset; and (c) a higher code rate (i.e., less redundant bits), which linearly increases the CPU utilization. 

The average MCSs assigned to the scheduled UEs in each RAN slice subnet must be considered by the RAN NSSMF to estimate the number of PRBs allocated for each RAN slice subnet in coarse time scales.\footnote{This is distinct from PRB scheduling, which allocates the PRBs assigned to each RAN slice subnet to their UEs} If the RAN NSSMF had a model of the vCPU consumption in function of the MCS and the PRBs, it could control the vCPU consumption in the shared DU instances by each RAN slice subnet. Using this model, the RAN NSSMF could implement mechanisms (e.g., for admission control) to avoid that RAN slice subnets exceed a given percentage of vCPU consumption for each shared DU instance, guaranteeing in this way their isolation at computing resource level.

Although the MCS and the number of PRBs are parameters controlled in the DU, they also impacts the vCPU consumption of a CU because the amount of user data processed by this gNB component directly depends on these parameters \cite{RRMslicing}. However, at the moment of writing this paper there are no models for the vCPU consumption in the CU because the works in the literature have not considered the CU/DU split and they have focused on the PHY layer (the most vCPU consuming).  

In addition to the vCPU consumption, the number of PRBs assigned to each RAN slice also impacts the performance of the vNICs used by the CU/DUs. This is due to more (less) PRBs involves more (less) user data processed by these vNICs. In the case that one of these gNB components is shared, the amount of processed user data by a shared vNIC depends of the PRBs assigned to each RAN slice subnet. By using a model that relates the number of PRBs with the mean waiting time of packets in the vNIC buffer, the RAN NSSMF could also control the PRB assignment for each RAN slice subnet in coarse time scales to avoid excessive buffer delays.

Models for the vCPU consumption and the waiting time on a vNIC buffer should be proposed in future researches to guarantee isolation between RAN slice subnets when a gNB component is shared.

\section{3GPP/NFV-based description model to manage the lifecycle of a shared gNB component}\label{sec:SharedManagementTemplates}
In \cite{OscarRANSlicing}, we proposed a model to describe the lifecycle management of the gNBs for several RAN slice subnets using the 3GPP/NFV management templates. Despite this model enables the customization of the gNBs and their adaptation to the temporal and spatial traffic demands of each RAN slice subnet, it assumes that the entire gNB is slice-specific (i.e., scenario \#1). In this work, we go a step further by proposing a description model that considers gNB sharing. 

Hereinafter, we discuss those aspects of the proposed description model that enables the sharing of a gNB component. For more detailed information about other aspects (i.e., regardless sharing), see \cite{OscarRANSlicing}. For clarity, we focus on scenario \#4. Notwithstanding, the proposed model can be easily adapted for scenarios \#2 and \#3.

Fig. \ref{fig:ProposalTemplates} shows the proposed model. The 3GPP/NFV management templates are hierarchically structured. On the left are the RAN NSSTs, each defining the FCAPS parameters and the gNBs of a RAN slice subnet. While the FCAPSs parameters are included in the RAN NSST, each gNB is described in a gNB Network Service Descriptor (NSD) that is referenced by the RAN NSST. Note that each RAN NSST references a specific gNB NSD. Managed by the NFVO, a gNB NSD contains a set of attributes to define the lifecycle management of the entire gNB. In this paper, we focus on the Scale Levels (SLs), Scaling Aspects (SAs), and Instantiation Levels (ILs) since they are key for the instantiation and scaling operations \cite{IFA014}. Each SL defines the number of CU (shared DU) instances and their resource capacity to guarantee the performance of the RAN slice subnet, given a specific traffic demand on a particular geographical area (e.g., a cellular infrastructure with 20 RUs). In turn, the SLs are grouped into two SAs. Each SA defines an independent scaling for the CU (or the shared DU) instances. To ease the management of the SLs of both SAs, the gNB NSD defines ILs. Each IL is the combination of two SLs, one per each SA (e.g, the IL \#2 is the combination of the SL \#1 for SA \#1 and SL \#2 for SA \#2).

To define the underlying virtual resources of the CU (shared DU) instances in each SL, the gNB NSD must reference a VNF Descriptor (VNFD) per each gNB component. Managed by the VNFM, a CU (shared DU) VNFD defines SLs, SAs, and ILs in a similar way as the gNB NSD. The main difference lies in the fact that these attributes directly define the VMs and their capabilities (i.e., number of vCPUs, CPU freq., etc) to accommodate the CU (shared DU) instances. In this description model, since each CU (shared DU) instance is mapped to a single VM, only one SA is required, thus the SLs and the ILs are used interchangeably in the CU (shared DU) VNFDs. Note that the shared DU VNFD is referenced by the gNB NSD of each RAN slice subnet while the CU VNFD is specific for each one.

Lastly, in addition to CU/DU VNFDs, the gNB NSD also references RU PNF Descriptors (PNFDs). Managed by the NFVO, each RU PNFD defines the physical connectivity points of a single RU. 

Deepening on the SAs of the gNB NSD, the SLs of SA \#1  define one CU instance whose resource capacity is described in the CU-ILs. Since a gNB has a unique CU, the SLs of SA \#1 are equivalent to the CU-ILs. Regarding SA \#2, each SL defines the required number of shared DU instances per each DU-IL (i.e., a VM with fixed capabilities).

When the DU instances are shared between several RAN slice subnets, the vCPU consumption in a DU instance might not affect to the majority of slice-specific CU instances. For example, if this increase is due to the user data of a single RAN slice subnet, only the vCPU consumption of one CU instance also increases. In this case, if a single SA was used in the gNB NSD, the design of the SLs would be more complex. Specifically, this design should consider (a) the number of RAN slice subnets that could share each DU instance; (b) their traffic demands; and (c) all the possible combinations for correlating the traffic demands on a shared DU and the slice-specific CUs.

The design complexity of using a single SA can be easily reduced if the shared DUs and the slice-specific CU are scaled independently. For that reason, two SAs have been defined in the gNB NSD, each for scaling independently the slice-specific CU of each RAN slice subnet, and the shared DUs. 

Defining two SAs is required but insufficient to scale shared DU instances. Since the gNB instances (i.e., including CU and DU instances) cannot be shared for all the RAN slice subnets, gNB instances per each RAN slice subnet should reference the same shared DU instances. In this case, if a shared DU instance needs to scale, multiple scaling operations should be triggered, one per gNB instance (and per RAN slice subnet). Furthermore, these scaling operations should be coordinated to select the same SL of SA \#2 to scale the DU. To avoid this scaling complexity, ETSI-NFV suggests the definition of an Auxiliary NSD \cite{IFA013}. This management template only defines ILs which coincide with the SLs for SA \#2. With this approach, when a shared DU needs to scale, the scaling operation is only executed in the auxiliary network service. After finishing this operation, the IL of each gNB instance must be updated according to the IL of the auxiliary network service. To that end, the SL of SA \#2 is equivalent to the new IL in the auxiliary netowrk service, and the SL of SA \#1 is changed according to the specific traffic demand of the RAN slice subnet that the CU belongs.

\section{Conclusions}\label{sec:Conclusions}
In this article, we shed light on the key aspects for sharing gNB components between RAN slice subnets. If the RRM algorithms at intra-slice level and the RRC layer are slice-specific or slice-aware, the gNB components could be shared because the treatment of the NR functionalities for the DRBs of each RAN slice subnet could be specifically configured. We have also identified that controlling the number of PRBs allocated to each RAN slice subnet and the MCSs assigned to their UEs, the isolation between RAN slice subnets can be guaranteed in a gNB component implemented as VNF. Finally, we have proposed a description model to define the lifecycle management of shared gNB components using the 3GPP/NFV management templates.

\section*{Ackowledgment}
This work is partially supported by the Spanish Ministry of Economy and Competitiveness and the European Regional Development Fund (Project TEC2016-76795-C6-4-R), the Spanish Ministry of Education, Culture and Sport (FPU Grant 17/01844) and the Andalusian Knowledge Agency (project A-TIC-241-UGR18).

\bibliographystyle{ieeetr}
\bibliography{references}

\end{document}